# First principles studies of a Xe atom adsorbed on Nb(110) surface

S. Dag[1], M. Shaughnessy*[2,3,4], C. Y. Fong[2], X. D. Zhu[2], L. H. Yang[3]

[1]*Lawrence Berkeley National Laboratory, Berkeley, CA 94720, USA*
[2]*Department of Physics, University of California, Davis CA 95616-8677, USA*
[3]*H Division, Lawrence Livermore National Laboratory, Livermore, CA 94551, USA*
[4]*Materials Physics Group, Sandia National Laboratories, Livermore, CA 94551, USA*

*Corresponding author: Michael Shaughnessy, PO Box 969, Livermore, CA 94551, *mcshaug@sandia.gov*

**ABSTRACT**

We study adsorption sites of a single Xe adatom on Nb(110) surface using a density functional theory approach: The on-top site is the most favorable position for the adsorption. We compare the binding features of the present study to earlier studies of a Xe adatom on close-packed (111) surface of face-centered cubic metals. The different features are attributed through a microscopic picture to the less than half filled d-states in Nb.

## 1. Introduction

Adsorptions of rare gas atoms on metallic surfaces serve as prototypes for heteroepitaxies involving lattice constant mismatches or frustrations caused by differences in crystal symmetries appearing at the interface between adsorbed materials and host substrates [1-3]. They are also used as ideal systems for investigating phonons and static defects in a substrate, on frictional interactions of ad-layers, and for patterning templates of spatially modulated structures in order to explore properties of low dimensional systems, such as pinning the Fermi energy in two dimensional systems [4]. Due to their importance in material fabrication for devices [5], there exists a wealth of experimental work concerning the properties of rare gas atoms adsorbed onto host metallic substrates, in particular on the closed-packed (111) surfaces of transition metals (TMs) with nearly filled or completely filled d-shell, such as Au, and Pt [6].

Among the TMs, Nb has less than half-filled 4d-states and exhibits superconducting properties. It crystallizes in the body-centered cube (bcc) - a more open structure than the close-packed fcc structure favored by metals having nearly filled and completely filled d-states. Nb has been increasingly used in spintronic devices for various tunneling junctions [7] and spin filters using Andreev reflection [8]. The latter use of Nb requires the metal to be in the superconducting state. Anticipating future spintronic devices, such as spin tunneling junctions, fundamental knowledge about absorptions on Nb surfaces will be crucial for making excellent devices. As part of our preliminary investigation into this area we carried out experiments for Xe on Nb(110) surface in its normal phase and determined the diffusion barrier to be 57.0 meV [3].

Theoretically, up to this point most efforts have focused on studying Xe adsorbed on the closed packed (111) surface of face-centered cubic metals, such as Cu, Pd, and Pt, with a *filled* or *nearly filled* d-shell. An algorithm based on density functional theory (DFT) [9] within the local density approximation (LDA) for the exchange-correlation (XC) between electrons [10] has been applied to model one and two Xe atoms adsorbed on Pt(111) surface [11]. The adsorption energy is overestimated by 10% with respect to the measured value [12]. Bagus *et al.* [13] used a scheme combining the generalized gradient approximation (GGA) [14] with a second order perturbation treatment of the van der Waals (vdW) interaction, ($1/r^6$), (GGA-vdW) to investigate a cluster model of the Xe atom on the Cu(111) surface. The adsorption energy at the on-top site (directly above a metallic atom) is -95 meV compared to the experimental value of -190 meV [15]. They attributed the discrepancy to the use of a cluster model instead of a surface model. Their calculations indicate a dipole is formed by shifting charges from the Cu atom to the Xe atom. Both the LDA and GGA were used by Betancourt and Bird [16] to study Xe on Pt(111) surface with a supercell model having five layers of Pt, one monolayer of Xe and an equivalent to nine layers of vacuum. Their results using LDA agree better with experimental adsorption site energies [17] than do their GGA ones. The charge distribution between a Xe atom and the on-top Pt determined from the LDA shows significant deviation from the spherical form. Da Salva *et al.* [18] reported results of a Xe atom adsorbed on a number of the close-packed (111) surface of face-centered-cubic (fcc) metals, such as Cu, Pd, and Pt using a supercell of five to six layers of metallic slab and a vacuum of 18 Å. One-fourth and one-ninth coverage of Xe were considered sufficient to justify neglecting the Xe-Xe interaction. These authors found a Xe adatom prefers the on-top site for adsorption. The adsorption energies at the on-top site of the (111) surfaces calculated using the LDA are -277 (-190) meV on Cu, -367 (-320) meV on Pt and -453 (-360) meV on Pd, respectively, where the numbers in the parentheses are the experimental values [19]. The corresponding energies on Pd and Pt surfaces using the GGA are only -76 and -82 meV, respectively. These authors attributed the on-top site preference to charge polarization of both the Xe atom and underlying noble or transition metal (TM) atom as a consequence of the Pauli repulsion. Lazić *et al.* [20] reported the results of potentials between a Xe and the substrate using the GGA with an intra-fragment treatment of the vdW interaction for models with one monolayer of Xe on the Cu(111) and Pt(111) surfaces, respectively. With the vdW interaction, the minimum separation between the Xe and the on-top Cu (Pt) atom is 3.20 (3.10) Å, compared to the experimental minimum of 3.60 (3.40) Å. These results suggest that for a single Xe atom on fcc metals the LDA gives a more realistic binding than the GGA combined with a vdW interaction.

At this point, there is essentially no theoretical report for adsorbing a Xe atom on normal state Nb surfaces. Instead of pursuing the more exotic Xe on superconducting Nb surface, it would be worthwhile to understand adsorption of a single rare gas atom on a Nb surface and compare to the more studied fcc metals, with nearly filled or filled d-states and the close-packed (111) surface. An understanding of the roles of both the openness of the bcc crystal structure and the less than half filled d-states of a Nb atom would extend our basic knowledge of adsorptions and illuminate future device fabrications.

In this paper, we report a theoretical study of a single Xe atom adsorbed on the more open Nb (110) surface. We address two basic issues: (A) Given the lower packing density and consequent altered surface character of the bcc (110) surface compared to the (111) surface of a fcc metal, will a Xe atom still prefer an on-top site for adsorption? If not, what is the most favorable site? If the answer to (A) is yes, (B) given the less than half-filled ($4d^4 5s^1$) *4d*-shell of the Nb atom, will the character of the binding between a Xe atom and its nearest neighbor (nn) Nb atom differ from the corresponding cases on fcc metal surfaces formed by a filled or nearly filled d-shell TM elements?

**2. Methods**

To model a Xe atom on Nb(110) surface, we used a supercell (Fig. 1(a)) that consists of a seven-layer slab region of Nb atoms and a vacuum region, taking up 42% of the supercell height, located on-top of the slab. In Fig. 1(a), the dashed lines outline the supercell. The Nb atoms are shown as light green-blue spheres and the Xe atom located above the substrate is shown in brown. We have a total of 56 Nb atoms in the supercell. The [$\bar{1}10$] direction in Fig. 1(a) is the *x*-axis of the supercell and the *y*-axis is along the [001] direction of the bcc structure. The supercell sizes in the *x* and *y* directions are twice as large as those of a conventional bcc unit cell so that the Xe-Xe interaction is negligible. The z-axis is along the [110] direction of the bcc structure. The bird's-eye view of the supercell surface is shown in Fig. 1(b). The dashed lines outline the unit cell in the *x-y* plane. The larger purple spheres are Nb atoms in the top layer and the smaller light green-blue spheres indicate Nb atoms in the second layer. Each layer consists of 8 Nb atoms.

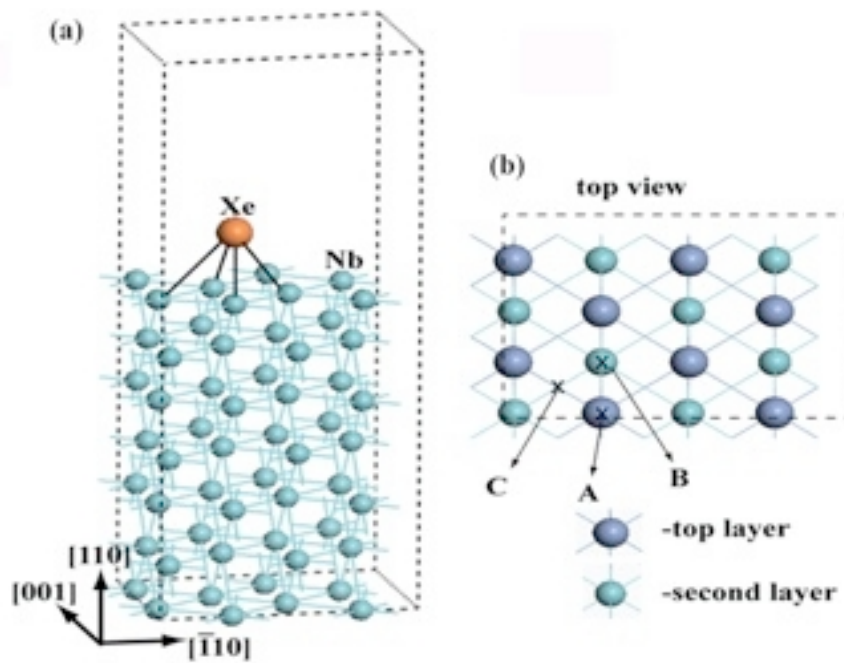

Fig. 1. (a) The supercell modeling the Nb(110) surface with a Xe atom on-top (left figure). The *x*-axis is along the [$\bar{1}$10] direction of the bcc cube, the *y*-axis is along [001] of the cube. (b) The top-view (right figure) of the supercell for Nb(110). Only the top two Nb layers are shown. **A** is the most favorable site, **B** is the hollow site, and **C** is the short-bridge site.

To compare the less than half-filled Nb d-states with nearly filled d-shell states of another TM, we chose to carry out the calculation of a Xe on Pd (111) surface. The fcc surface is modeled with six layers in the form of the so-called ABC packing (no relation to the sites shown in Fig.

1(b)) with the Xe located on-top of one of Pd atoms in the topmost layer of the slab. The vacuum region has the same width as the slab.

As shown by earlier calculations of Xe on fcc metals, the LDA gives more reasonable binding between a single adatom and the atoms at the surface. Furthermore, we determined the diffusion barrier to be 30 meV, which is 55.7% of the measured value. We checked and found that using a GGA-vdW functional gives the ontop site as the favorable site but there is a negligible barrier (0.03 meV) for the Xe atom on the slab. Therefore, in this paper we report the LDA to determine the most favorable site for a Xe atom adsorbed on Nb(110) surface and to illustrate the difference of the bonding features between the Xe on Nb(110) and Xe on Pd(111) surface. We believe the qualitative features are valid using the LDA electron-electron correlations and so we purposely chose not to include the GGA-vdW results.

We used the VASP algorithm [21] based on density functional theory [10] to compute the total energy at each site of a Xe atom on Nb(110) and examined the binding features of the Xe atom adsorbed at the most favorable site. The atoms are characterized by the PAW pseudopotentials determined with normal valence electronic configurations [22]. Planewaves were used as basis functions. The separation between the Nb atoms in the slab was determined by optimizing the lattice constant of a bcc Nb crystal. The value determined by the LDA is 3.27 Å. The measured value is 3.30 Å [23]. Relaxations were carried out to reduce the forces acting on the atoms due to the presence of the vacuum region and Xe adatom. A relaxation is complete when all the components of any force on any atom are less than or equal to 6.0 meV/Å. The kinetic energy cutoff for the planewave basis is set to 650.00 eV. The Monkhorst-Pack mesh [24] of 11×13×1 was used to generate the special **k**-points for constructing the charge density. The convergence of the total energy is better than 1.0 meV.

For Xe on Pd surface calculation, the optimized LDA lattice constant for Pd is 3.85 Å, as compared to the experimental value of 3.89 Å and the cutoff energy and **k**-point mesh are 300 eV and 10x10x1, respectively.

**3. Results and Discussion**

In the following, we address the two issues.

*3.1 Will a Xe atom still prefer an on-top site?*

In Fig. 1(b), there are three special sites, **A**, **B** and **C**, competing to most favorably bind the adatom. The **A** site, the on-top site, is where the Xe atom sits atop a Nb atom at the uppermost layer of the slab. The **B** site, the hollow site, is where the adatom rests directly above a second layer Nb atom and the **C** site, the bridge site, is between the **A** and **B** sites.

The **A** site shown in Fig. 1(b), is energetically the most favorable, with total energy -614.521 eV. The highest total energy is associated with the **B** site (Fig. 1(b)), -614.428 eV. The total energy at the **C** site falls in between at -614.491 eV. This result is qualitatively consistent with the studies of Xe on other TM metals. *The open structure of the Nb does not affect the on-top configuration of the adatom as the most favorable site found in close-packed fcc (111) surface.*

Using the total energy at the **A** site including the relaxation (-614.521 eV), the energy of the relaxed pure Nb (110) surface (-614.195 eV), and the calculated chemical potential of a single Xe atom in the same supercell (-0.042 eV), the adsorption energy at **A** is 284.0 meV, which is between the adsorption energies of the rare gas atom on Cu(111) and Pt(111) surfaces [18].

To compare to the experimental diffusion barrier, the potential surface for the Xe atom was calculated on a grid of 16 points in a portion of the surface unit-cell shown in Fig. 1(b). The results are shown in Fig. 2. The **C** site sits at a saddle point of an energetic obstacle for a Xe atom to move from one **A** site to a neighboring one. The calculated energy barrier to move a Xe atom along to path **A – B – A** is 30 meV. Compared to the experimental diffusion barrier value of 53.87 meV reported by Thomas *et al*. [3], the calculated barrier (energy difference between sites **A** and **C**, a saddle point) is only 55.7% of the experimental value. The discrepancy between the calculated barrier and the measured value may be partly due to the Xe-Xe attraction in experiments carried out with a finite coverage, which tends to raise the effective diffusion barrier, as discussed through differences in binding energies of a single Xe atom and a monolayer of Xe atoms on fcc metals.

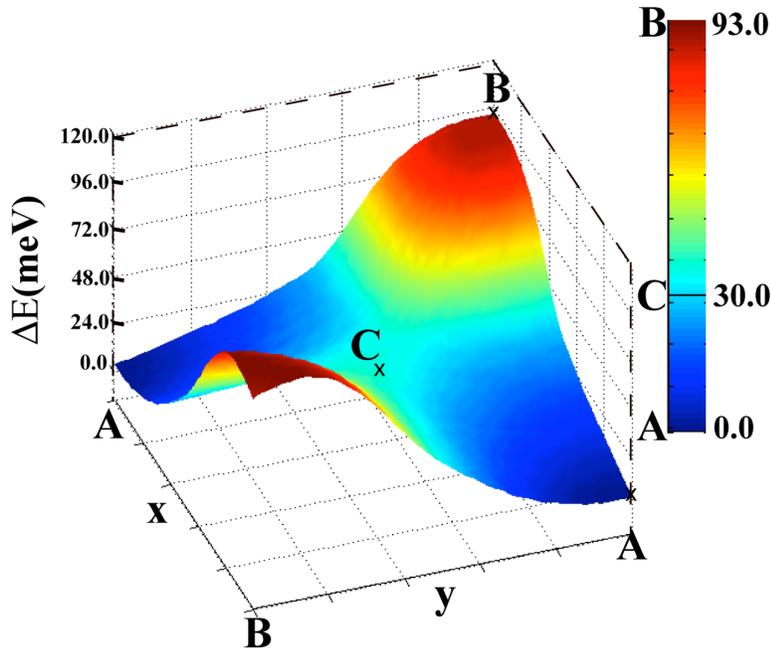

Fig. 2: (a) The potential energy of a Xe atom on Nb(110) is defined by Eq. (1) in meV, over the lower 1/16-th section of the surface unit cell, calculated with a 4x4 mesh in the plane. The energy difference between the **A** and the **C** site determines the barrier for diffusion.

### 3.2 *Binding characteristics at site A*

We now focus on the binding characteristics of the Xe atom at the **A**-site. The *total* valence charge distribution in the *x-z* plane (Fig. 1) containing Xe and its nn-Nb is shown in Figs. 3(a). Xe is located at 2.31 Å from the left edge of the supercell in the *x*-direction. In the *z*-direction, it is at 3.13Å above the nn-Nb, and is at 16.72Å from the bottom of the Nb slab. In order to see the complete charge distributions of the Nb atoms located at the bottom layer of the slab, the section shown in Fig. 3(a) is defined between -2.19 Å and 22.09 Å instead of 0 and 24.28 Å. The vacuum region can be clearly seen to ensure our supercell adequately characterizes the surface.

To clearly see the redistribution of charge density between the Xe atom and its nn-Nb, we computed the charge density difference in the same *x-z* plane defined as follows,

$$\Delta\rho = \rho_{tot} - \rho_{pure} - \rho_{Xe}, \tag{1}$$

where $\rho_{Xe}$ is the charge density of an isolated Xe atom placed at the same position as shown in Fig. 3(a) or Fig. 3(c) but without the Nb slab. $\rho_{pure}$ is the charge density in the supercell model with the same atomic configurations for $\rho_{tot}$ without the Xe atom. $\rho_{tot}$ is the charge density of the relaxed Nb slab with the Xe atom at the **A** site. The difference charge density, $\Delta\rho$, is shown in Figs. 3(b). We plot $\Delta\rho$ from z = 11.45 to z = 22.09 Å, a partial section of the ones in Figs. 3(a).

The redistribution shown in Fig. 3(b) exhibits two features: (a) The most prominent feature of $\Delta\rho$ is a net charge accumulation shown in the red region localized between the two atoms, and (b) there are the two upward pointing lobes (yellow regions marked by contour value of 0.012) just above the nn-Nb and streaks (green-yellow regions) just below the nn-Nb atom. To clearly show the character of feature (a), we plot $\Delta\rho$ in Fig. 3(c) along the line joining the Xe and nn-Nb atoms. The positive peak corresponds to the red region in Fig. 3(b) with the maximum value of 0.032 e/Å$^3$. The distance between the center of the nn-Nb and the maximum is 1.75 Å, which is larger than both the covalent and atomic radii of Nb, 1.34 and 1.46 Å, respectively [26]. The separation between the center of the Xe and the center of the red region is 1.57 Å, which is also larger than 1.31 Å, the covalent radius of the Xe atom [26]. Therefore, both atoms contribute their charge densities to this red region. To conserve the charge, there are two primary charge depletion regions: (i) Behind the Xe atom (deep blue region), and (ii) immediately above and below the nn-Nb atom (deep blue regions). These results suggest the $p_z$-state of the Xe atom and $d_{z^2}$-state of the nn-Nb are the orbitals being redistributed during the adsorption. From feature (b), the $d_{xz}$-state of the Nb is also a part of the redistribution.

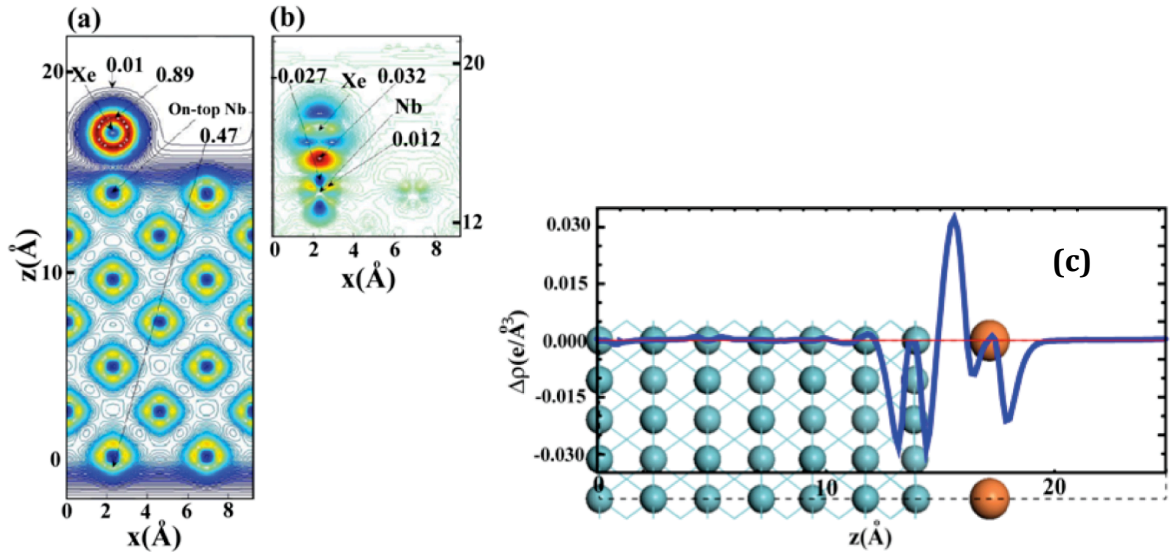

Fig. 3. The LDA results: (a) Contour plot of the *total* valence charge density in the *x-z* plane (Fig. 1) passing through Xe and its nn Nb atoms. (b) Contour plot of the difference charge density computed from Eq. (1) in the important partial section of the same plane shown in (a). (c) The outline of the LDA difference charge density along the line passing through the two atoms.

To see whether similar features are exhibited for the Xe atom on an fcc metal surface, we show in Fig. 4(a) and (b) charge distribution and the difference charge density (Eq. (1)) for a Xe on Pd (111) surface. The Xe atom is at an on-top site, 8.18Å along the vertical axis. The on-top Pd is located at 5.33Å. The separation of the two atoms is 2.85Å and is about 0.5Å larger than the case Xe on Nb(110). There are three regions where the densities increase, namely close to the center of the Xe atom, the middle region between the adatom and its nn, which is the region with the greatest increase, and the regions to the right and left of the Pd atom. The most charge-depleted region due to the presence of the Xe is just above and below the Pd atom. These results qualitatively agree with the results given in Ref. 16.

Comparing the results given in Fig. 4(b) to Fig. 3(b), both figures show accumulations of charge between the adatom and its nn metal atom. The maximum value (0.017 e/Å$^3$) in Fig. 4(b) is at 1.49 Å above the Pd atom and 1.36 Å below the Xe atom. The atomic and covalent radii of Pd are 1.37 and 1.28 Å [25]. This maximum is just outside of both the Xe (having the covalent radius of 1.31 Å) and the Pd atom. The corresponding value in Fig. 3(b) is 0.032 e/Å$^3$, which is almost a factor 2 larger. We attribute this difference to the less than half filled d-states in Nb and the nearly filled ones in Pd. The results of Xe on Cu (111) surface [16] give further evidence of less accumulation in this region due to the filled d-states in Cu. Fig. 6 in Ref. 16 clearly shows the charge redistribution in this middle region of the Xe/Cu system is significantly less than the ones for Xe/Pd. With the trend from Nb, Pd to Cu, the accumulation of charge density in the region between the Xe and the nn metal atom decreases as the occupation of the d-states increases.

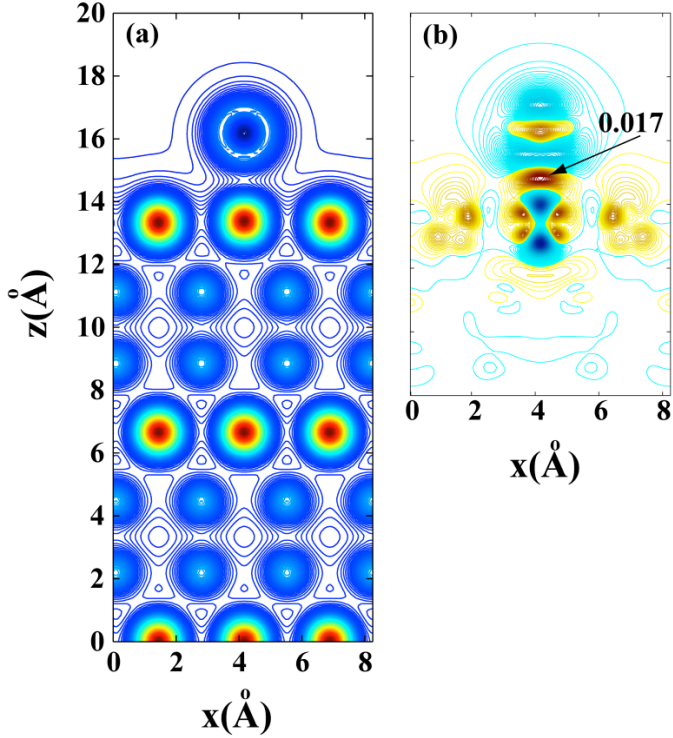

Fig. 4. (a) Contour plot of the *total* valence charge density in the plane passing through Xe and its nn Pd atoms. (b) Contour plot of the charge density difference computed from Eq. (2) in the important partial section of the same plane as shown in (a).

We now comment on the physical origin of the accumulation in Xe on Nb(110). The $d_{xz}$ orbital of the on-top Nb is the least occupied before the Xe is adsorbed, as shown in Fig. 3(a), in which there are no lobes showing the $d_{xz}$ character in the bulk or near the surface. Upon Xe adsorption, the $d_{xz}$ orbital gains some charge from the Xe $p_x$-state owing to the close proximity of the two states. However, the Xe adatom prefers a closed shell electronic configuration. To compensate for the loss of the $p_x$-electron, it shifts the charge from its $p_z$ state. This $p_z$ state makes up its loss by pulling the charge of the $d_{z^2}$ state of the on-top Nb. The tightly bound character of the $d_{z^2}$ state contributes its charge to the red region shown in Fig. 3(b). This explains not only why the largest charge redistribution region (the red region) is outside the atomic radii of both atoms but also the increases of charge distributions shown in light yellow at the sides of the Xe atom and just above the Nb atom. The overall redistribution is not solely from the $p_z$ and the $d_{z^2}$ states due to Pauli repulsion. It involves other states. This description can also explain the decreased accumulation when the d-states are more occupied. When the $d_{xz}$ state is filled, the nearby $p_x$ electron is less affected by the d-shell of the TM. Consequently, the occupancy of the $p_z$ state is less affected. So is the $d_{z^2}$ state of the TM element. This is shown in the Xe/Cu case.

## 4. Conclusions

We determined the most favorable location of a Xe atom adsorbed on a Nb(110) surface using the LDA. Therefore, the results are qualitatively consistent in determining the most favorable site of adsorption for a single Xe atom on other TM surfaces. *The different crystal structures of the Nb and Pt or Pd surfaces do not play an important role.* The **B** site has a higher energy than the **C** site. The diffusion barrier is 30.0 meV, which is only 55.7% of the measured value. The discrepancy can be due to the fact that an isolated Xe atom on the surface is considered.

By plotting the difference of the charge densities (Eq. (1)) and comparing the cases of the Xe atom on Nb and Pd, a common binding feature of the Xe atom at the A site is shown to be a charge accumulation between the Xe atom at the on-top site and its nn TM atom underneath. The amount of accumulation for different TM elements can be attributed to the filling of the d-orbitals of the TM atom. A microscopic mechanism is suggested to explain the detailed features of the charge redistributions during the adsorption. The differing charge redistributions for a Xe atom on Nb(110) and Pd(111) due to the filling of the d-states may inform the design of growth methods for delicate spintronic devices.

## Acknowledgments

This work was supported in part by NSF under NSF-ECCS-0725902, by the San Diego Supercomputer Center, and by the donors of Petroleum Research Fund, administered by the American Chemical Society. Work by LHY was performed under the auspices of the U. S. Department of Energy by Lawrence Livermore National Laboratory under contract No. DE-AC52-07NA27344. SD is supported by DOE under grant No. DE-AC02-05-CH11231.

**Captions:**

Fig. 1. (a) The supercell modeling the Nb(110) surface with a Xe atom on-top (left figure). The *x*-axis is along the [$\bar{1}$10] direction of the bcc cube, the *y*-axis is along [001] of the cube. (b) The top-view (right figure) of the supercell for Nb(110). Only the top two Nb layers are shown. **A** is the most favorable site, **B** is the hollow site, and **C** is the short-bridge site.

Fig. 2: (a) The potential energy of a Xe atom on Nb(110) is defined by Eq. (1) in meV, over the lower 1/16-th section of the surface unit cell, calculated with a 4x4 mesh in the plane. The energy difference between the **A** and the **C** site determines the barrier for diffusion.

Fig. 3. The LDA results: (a) Contour plot of the *total* valence charge density in the *x-z* plane (Fig. 1) passing through Xe and its nn Nb atoms. (b) Contour plot of the difference charge density computed from Eq. (1) in the important partial section of the same plane shown in (a). (c) The outline of the LDA difference charge density along the line passing through the two atoms.

Fig. 4. (a) Contour plot of the *total* valence charge density in the plane passing through Xe and its nn Pd atoms. (b) Contour plot of the charge density difference computed from Eq. (2) in the important partial section of the same plane as shown in (a).